\documentclass[aps,prb,twocolumn,showpacs,showkeys,superscriptaddress]{revtex4-1}

\usepackage{graphicx}
\usepackage{dcolumn}
\usepackage{bm}
\usepackage{multirow}
\usepackage{amsmath,amssymb}
\usepackage [latin1]{inputenc}

\begin{document}

\preprint{}

\title{Metamagnetic transitions and anomalous magnetoresistance in EuAg$_{4}$As$_{2}$ crystal}


\author{Qinqing Zhu}
\thanks{These authors contributed equally to this work.}
\affiliation {Hangzhou Key Laboratory of Quantum Matter, Department of Physics, Hangzhou Normal University, Hangzhou 311121, China}

\author{Liang Li}
\thanks{These authors contributed equally to this work.}
\affiliation {Hangzhou Key Laboratory of Quantum Matter, Department of Physics, Hangzhou Normal University, Hangzhou 311121, China}

\author{Zhihua Yang}
\affiliation {Hangzhou Key Laboratory of Quantum Matter, Department of Physics, Hangzhou Normal University, Hangzhou 311121, China}

\author{Zhefeng Lou}
\affiliation {Department of Physics, Zhejiang University, Hangzhou 310027, China}

\author{Jianhua Du}
\affiliation {Department of Physics, China Jiliang University, Hangzhou 310018, China}

\author{Jinhu Yang}
\affiliation {Hangzhou Key Laboratory of Quantum Matter, Department of Physics, Hangzhou Normal University, Hangzhou 311121, China}

\author{Bin Chen}
\affiliation {Hangzhou Key Laboratory of Quantum Matter, Department of Physics, Hangzhou Normal University, Hangzhou 311121, China}

\author{Hangdong Wang}
\email{hdwang@hznu.edu.cn}
\affiliation {Hangzhou Key Laboratory of Quantum Matter, Department of Physics, Hangzhou Normal University, Hangzhou 311121, China}

\author{Minghu Fang}
\email{mhfang@zju.edu.cn}
\affiliation {Department of Physics, Zhejiang University, Hangzhou 310027, China}
\affiliation {Collaborative Innovation Center of Advanced Microstructures,  Nanjing University, Nanjing 210093, China}



\date{\today}

\begin{abstract}
In this paper, the magnetic and transport properties were systematically studied for EuAg$_{4}$As$_{2}$ single crystals, crystallizing in a centrosymmetric trigonal CaCu$_{4}$P$_{2}$ type structure. It was found that two magnetic transitions occur at $\textit{T}$$_{N1}$ = 10 K and $\textit{T}$$_{N2}$ = 15 K, respectively, which are driven to lower temperatures by applied magnetic field. Below $\textit{T}$$_{N1}$, two successive metamagnetic transitions were observed when a magnetic field is applied in the $\textit{ab}$ plane ($\textit{H}$ $\parallel$ $\textit{ab}$ plane). For both $\textit{H}$ $\parallel$ $\textit{ab}$ and $\textit{H}$ $\parallel$ $\textit{c}$, EuAg$_{4}$As$_{2}$ shows a positive, unexpected large magnetoresistance (up to 202\%) at low fields below $\textit{T}$$_{N1}$, and a large negative magnetoresistance (up to -78\%) at high fields/intermediate temperatures, which may have potential application in the magnetic sensors. Finally, the magnetic phase diagrams of EuAg$_{4}$As$_{2}$ were constructed for both $\textit{H}$ $\parallel$ $\textit{ab}$ and $\textit{H}$ $\parallel$ $\textit{c}$ by using the resistivity and magnetization data.

\end{abstract}


\maketitle
\section{INTRODUCTION}

Responses of Eu-based compounds to external fields have generated immense interest due to exhibiting many exotic properties, such as valence transition \cite{A.Mitsuda}, Kondo behavior \cite{C.Feng}, quantum hall effect \cite{H.Masuda}, novel magnetoresistance (MR) \cite{S.Majumdar, C.Yi}, resulted from their complicated, tunable magnetic ground states with large local moments. In those modulated systems with an incommensurate magnetic structure induced by the strong competing interactions involving the magnetic moments, the lattice, and/or the conduction electrons, like the long-range
Ruderman- Kittel- Kasuya- Yosida (RKKY) type coupling, one or several successive transitions have been observed at low temperatures when a magnetic field is applied, leading to complex magnetic phase diagrams \cite{D.Gignoux}.

The Eu-based ternary pnictide EuAg$_{4}$As$_{2}$, crystallizing in a centrosymmetric trigonal CaCu$_{4}$P$_{2}$ type structure (space group $\textit{R}$$\bar{3}$$\textit{m}$, No. 166), was first reported by Stoyko $\textit{et al}$ \cite{S.Stoyko}. The structure can be considered as the derivative from the trigonal CaAl$_{2}$Si$_{2}$-type structure, by inserting an additional itinerant Ag$_{2}$ layer between the close-packed Ag$_{2}$As$_{2}$ layers. Unlike in CaCu$_{4}$P$_{2}$, where the Cu2 sites are fully occupied, the Ag2 sites are split into three isotropic and equally partially occupied sites in EuAg$_{4}$As$_{2}$ [see Fig. 1(b)]. To the authors' knowledge, a few studies on the physical properties of the ternary CaCu$_{4}$P$_{2}$ type pnictides AAg$_{4}$Pn$_{2}$ (A = Sr, Eu; Pn = As, Sb) have been reported. For SrAg$_{4}$As$_{2}$, the quantum oscillation measurements reveal small Fermi pockets with light effective masses and unexpected high mobilities, in contrast with the predication of the first-principle calculations \cite{B.Shen}. For the magnetic EuAg$_{4}$As$_{2}$, based on the measurements of magnetization \cite{B.Gerke}, neutron diffraction \cite{B.Shen2}, $^{151}$Eu M\"{o}ssbauer spectroscopy \cite{B.Gerke, D.H.Ryan} and pressure effects \cite{L.Sergey}, it was found that a structural distortion occurs at about 120 K, and two magnetic transitions emerge at 15 K and 9K, respectively, below 9 K, an incommensurate, non-collinear long-range antiferromagnetic (AFM) state; between 9 - 15 K, a long-range magnetic order with the Eu$^{2+}$ 4$f^7$ spin moments having an incommensurate sine modulated structure. However, there are no reports about the magnetotransports and the phase diagram for this low-dimensional system with so rich magnetic structures.

In this article, we report the magnetization and MR measurements on EuAg$_{4}$As$_{2}$ single crystal. It is confirmed that two magnetic transitions occur at $\textit{T}$$_{N1}$ = 10 K and $\textit{T}$$_{N2}$ = 15 K, respectively, with the magnetic moments almost lying in the $\textit{ab}$ plane. We further observed that the magnetic transition temperatures, $T_{N1}$ and $T_{N2}$, decrease with increasing magnetic field and two successive metamagnetic (MM) transitions occurring at 0.5 T and 0.95 T field applied in the $ab$ plane at 2 K. Interestingly, it was found that an anomalous MR emerges for both $\textit{H}$ $\parallel$ $\textit{ab}$ and $\textit{H}$ $\parallel$ $\textit{c}$ orientations, with different magnetic field dependence at various temperature range, which is related to the magnetic ground states. Finally, we constructed the phase diagrams based on the data of the magnetization and resistivity measurements.

\section{EXPERIMENTAL METHODS}
\begin{figure}
  \includegraphics[width=8cm]{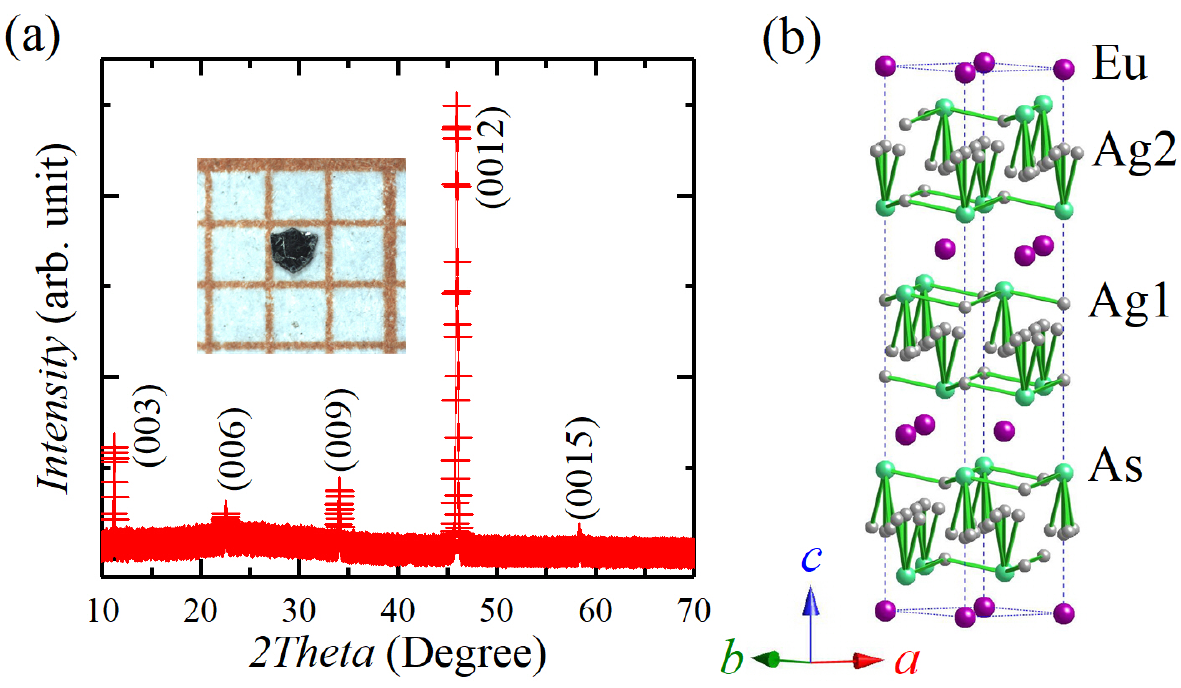}\\
  \caption{(Color online) (a) Single crystal XRD pattern of EuAg$_{4}$As$_{2}$. Inset: A photograph of a EuAg$_{4}$As$_{2}$ single crystal. (b) Crystal structure of EuAg$_{4}$As$_{2}$. Europium, silver, and arsenic atoms are drawn as purple, grey, and green spheres, respectively.}
\end{figure}

EuAg$_{4}$As$_{2}$ crystals were grown using a self-flux method. First, Eu chunks, and Ag, As powders were mixed with a ratio of 1:4:2 and were put into an alumina crucible and sealed in an evacuated silica tube. The mixture was heated up to 1100 $^{\circ}$C and kept for 24 hours, then cooled down to 700 $^{\circ}$C at a rate of 3 $^{\circ}$C/h. Finally the furnace was cooled to room temperature after shutting down the power. Single crystals with a typical dimension of 0.8$\times$0.8$\times$0.2 $ mm^3$ [see the inset of Fig. 1(a)], were mechanically exfoliated from the flux. The crystal composition was determined by the Energy-dispersive X-ray spectroscopy (EDX) in a Zeiss Supra 55 scanning electron microscope to be the stoichiometric EuAg$_{4}$As$_{2}$. The crystal structure was confirmed by X-ray diffraction (XRD) performed at room temperature on a Rigaku X-ray diffractometer with Cu $\textit{K}$$\alpha$1 radiation [see Fig. 1(a)]. All XRD peaks are indexed to be (00$\textit{l}$) planes, and the cell parameter $\textit{c}$ is yielded to be about 23.65 ${\AA}$, in consistent with the previous result \cite{B.Shen2}. The magnetization was measured using the \textit{Quantum Design} Magnetic Properties Measurement System (MPMS-VSM-7T). The resistivity measurements were carried out by the standard four-probe technique on the \textit{Quantum Design} Physical Properties Measurement System (PPMS-9T).

\section{RESULTS AND DISCUSSIONS}

\begin{figure}
  \includegraphics[width=8cm]{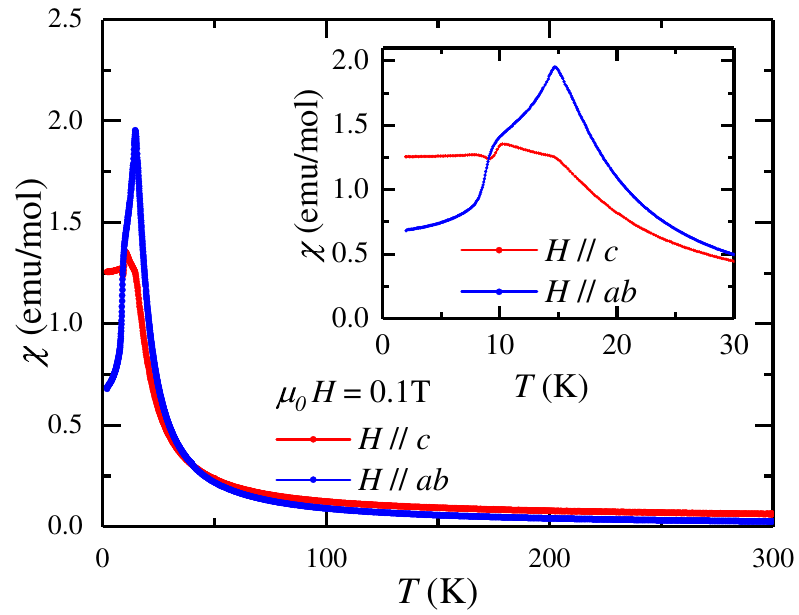}\\
  \caption{(Color online) Temperature dependence of the magnetic susceptibility for EuAg$_{4}$As$_{2}$ measured at $\mu_0H$ = 0.1 T applied in the $\textit{ab}$ plane and along the $\textit{c}$ axis, respectively. The inset of the figure shows an enlarged view of magnetic susceptibility at low temperatures.}
\end{figure}

\begin{table}
\caption{\label{Table1}%
The fitting parameters of the high temperature susceptibility for EuAg$_{4}$As$_{2}$ single crystal.
}
\begin{ruledtabular}
\begin{tabular}{ccc}
\textrm{Parameters}&
\textrm{\textit{H} $\parallel$ \textit{c} axis}&
\textrm{\textit{H} $\parallel$ \textit{ab} plane}\\
\colrule
$\chi_{0}$ (emu/mol) & 0.03578 & -0.00228\\
$\textit{C}$ (emu$\cdot$K/mol) & 7.88 & 7.91 \\
$\theta$ (K) & 10.3 & 14.1\\
$\mu_{eff}$ ($\mu_{B}$/$\textit{Eu}$) & 7.94 & 7.95\\
\end{tabular}
\end{ruledtabular}
\end{table}

Figure 2 shows the temperature dependence of magnetic susceptibility, $\chi$$_{ab}$ ($\textit{H}$ $\parallel$ \textit{ab} plane) and $\chi$$_{c}$ ($\textit{H}$ $\parallel$ $\textit{c}$ axis), respectively, measured at magnetic field of 1 kOe with a zero field cooling (ZFC) process for a EuAg$_{4}$As$_{2}$ crystal. The higher temperature ($>$ 50 K) $\chi$$_{ab}$($\textit{T}$) data can be well fitted by the Curie-Weiss law, $\chi=\chi_{0}+ \frac{C}{T-\theta}$, where $\chi_{0}$ and $\textit{C}$ is the temperature-independent constants, $\theta$ is the Curie temperature. The fitting result yields an effective moment $\mu$$_{eff}$ = 7.95 $\mu$$_{B}$, close to the theoretical value of Eu$^{2+}$ moments ($g\sqrt{S(S+1)}$ = 7.94 $\mu$$_{B}$, \textit{S} = 7/2 and \textit{g} = 2), implying that Eu$^{2+}$ 4$f^7$ electrons are localized, and a positive Curie temperature $\theta$ = 14.1 K, indicating ferromagnetic (FM) interactions in the paramagnetic (PM) regions (see Table I). At low temperatures, $\chi$$_{ab}$(\textit{T}) exhibits a sharp peak at about 15K ($\textit{T}$$_{N2}$) and a kink at 10K ($\textit{T}$$_{N1}$), where the transition temperature $\textit{T}$$_{N1}$ is a little bit higher than that reported previously \cite{B.Shen2} due to the smaller field applied in our measurement, as discussed below. The result confirms that the triangle Eu$^{2+}$ spin sublattices undergo a transition from a PM state to an incommensurate sine modulated AFM state (referred to as AFM-II) at $\textit{T}$$_{N2}$ and then to an incommensurate, non-collinear AFM state (referred to as AFM-I) at $\textit{T}$$_{N1}$ \cite{B.Shen2, D.H.Ryan}. The $\chi$$_{c}$($\textit{T}$) has a similar behavior to that in $\chi$$_{ab}$($\textit{T}$) at high temperatures, but $\chi$$_{c}$($\textit{T}$) remains almost unchanged and is larger than that of $\chi$$_{ab}$($\textit{T}$) below $T_{N1}$, suggesting that the Eu$^{2+}$ moments almost lie in the $\textit{ab}$ plane. Besides, it should be noted that the large residual magnetic moment at $\textit{T}$ = 2 K is consistent with the incommensurate, non-collinear AFM ordered state.

\begin{figure*}
  \includegraphics[width=16cm]{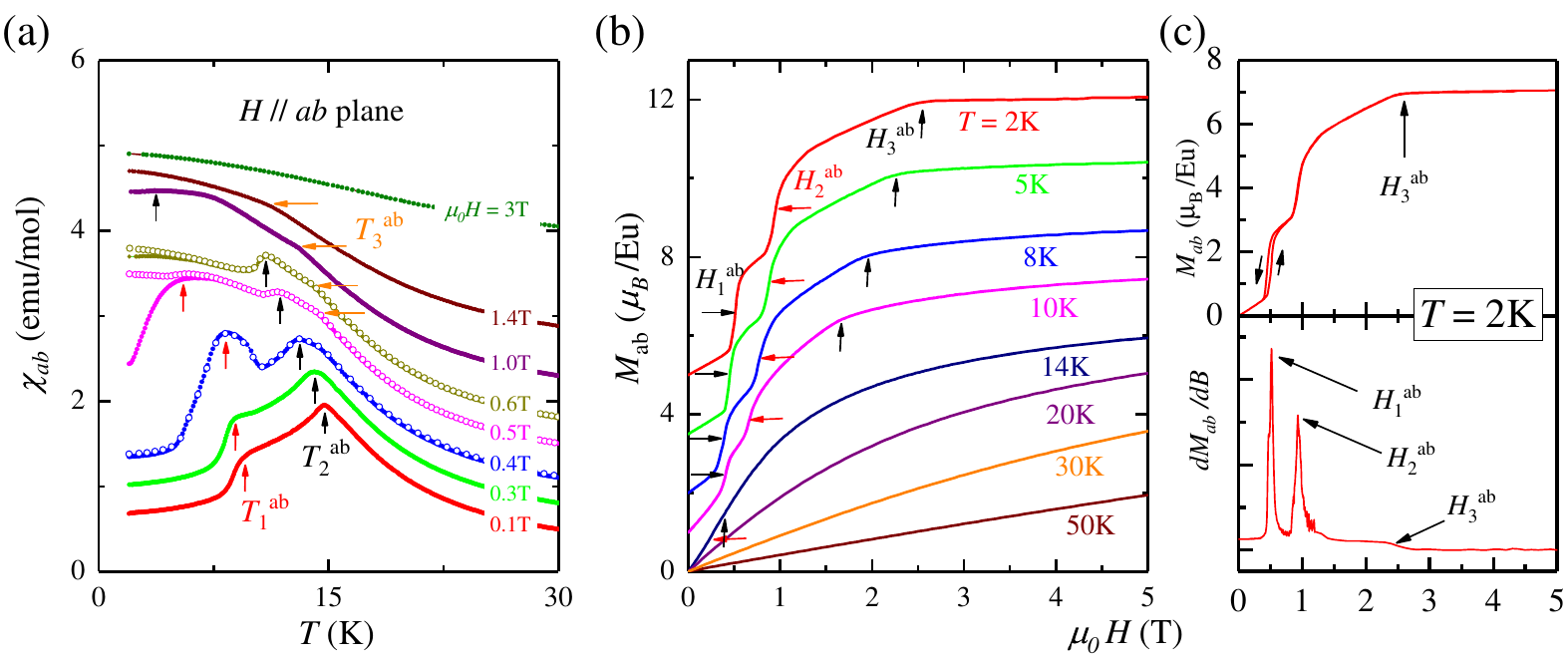}\\
  \includegraphics[width=16cm]{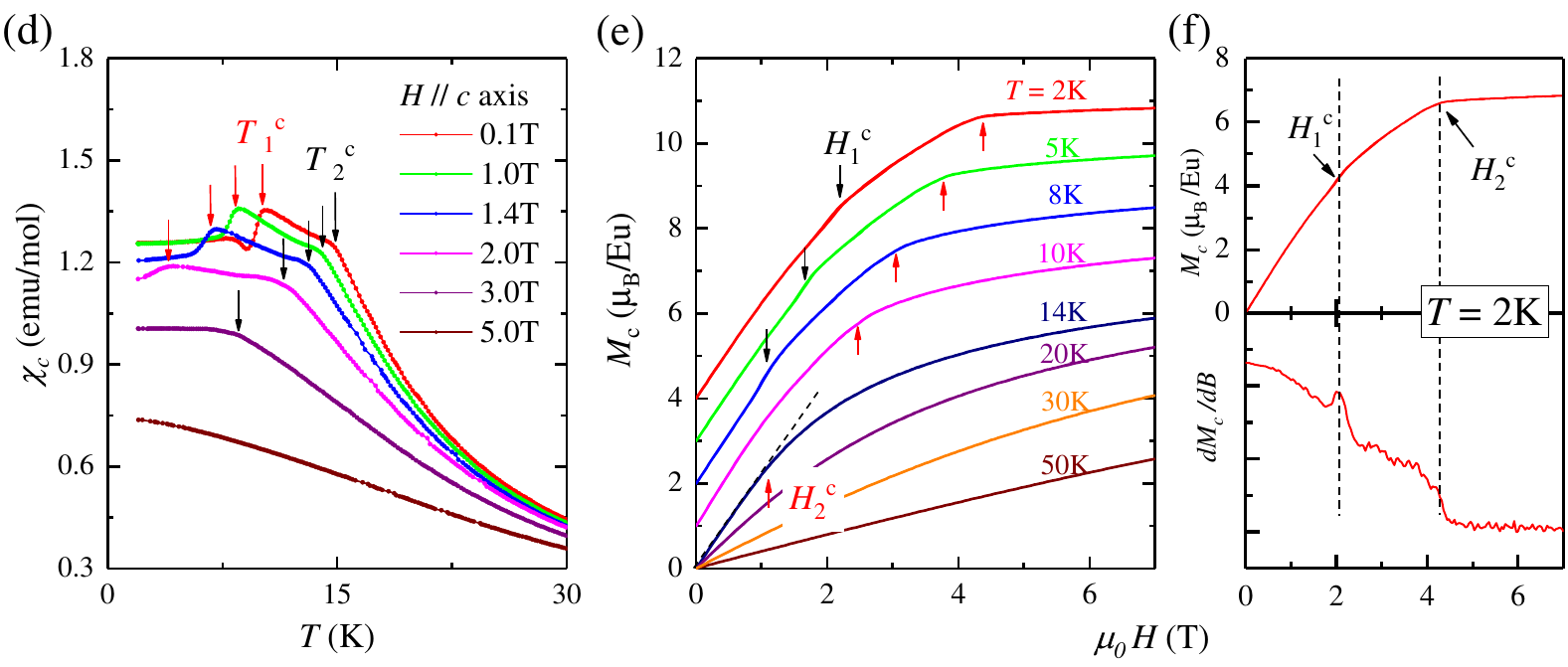}\\
  \caption{(Color online) Temperature dependence of magnetic susceptibility of the EuAg$_{4}$As$_{2}$ single crystal under several selected magnetic fields for $\textit{H}$ $\parallel$ $\textit{ab}$ (a) and $\textit{H}$ $\parallel$ $\textit{c}$ (d), below $\textit{T}$ = 50 K. For some selected magnetic fields along the $\textit{ab}$ plane, both field-cooled (FC, hollow circles) and zero-field-cooled (ZFC, solid circles) data are shown. For most other cases, only the ZFC data are shown. In the case of $\textit{H}$ $\parallel$ $\textit{ab}$, the data have been shifted for clarity except for the one taken in the field of 0.1 T. Field dependence of magnetization of the EuAg$_{4}$As$_{2}$ single crystal at various temperatures, with the field parallel to the $\textit{ab}$ plane (b) and $\textit{c}$ axis (e). The data taken below $\textit{T}$ = 14 K has been shifted for clarity. (c) The field dependence of $\textit{M}$$_{ab}$ (top) and its first derivative (bottom) taken in the field increasing process at $\textit{T}$ = 2 K. (f) The field dependence of $\textit{M}$$_{c}$ (top) and its first derivative (bottom) at $\textit{T}$ = 2 K. The dashed lines are guides to the eyes.}
\end{figure*}

Figure 3(a) shows $\chi$$_{ab}$(\textit{T}) measured at several magnetic fields. At $\mu_{0}$$\textit{H}$ = 0.1 T, the AFM-II transition occurs at $\textit{T}$$_{2}$$^{ab}$ = 15 K and the AFM-I transition occurs at $\textit{T}$$_{1}$$^{ab}$ = 10 K, as discussed above. It can be seen that with increasing field, both $T_1$$^{ab}$ and $T_2$$^{ab}$ are noticeably shifted to lower temperatures. Meanwhile, an additional transition is observed at $\textit{T}$$_{3}$$^{ab}$ under an external field, characterized by a tiny kink in the $\chi$$_{ab}$($\textit{T}$) curves. For 0.3 T $\leq$ $\mu_{0}$$\textit{H}$ $\leq$ 0.6 T, we observe a clearly deviation between ZFC and FC curves at low temperatures, which may be due to the MM transitions occurring in this field range as discussed below. From the $\chi$$_{c}$(\textit{T}), it can also be seen that both $T_1$$^{c}$ and $T_2$$^{c}$ are shifted to lower temperature with increasing magnetic field as shown in Fig. 3(d).

In order to understand these peculiar behaviors of $\chi$($\textit{T}$), we carried carefully out the isothermal magnetization measurements. Figure 3(b) shows the field dependence of magnetization, $M_{ab}(H)$, for $\textit{H}$ $\parallel$ $\textit{ab}$ up to 5 T. At $\textit{T}$ = 50 K, the $\textit{M}$$_{ab}$ increases nearly linearly with increasing field, indicating a typical paramagnetic behavior. When 14 K $\leq$ $T$ $\leq$ 30 K, the $M_{ab}(H)$ curves exhibit an apparent nonlinear behavior due to the magnetic fluctuation close to the $T_{N2}$. At $\textit{T}$ $\leq$ 10 K, the $M_{ab}(H)$ exhibits two jumps at $\textit{H}$$_{1}$$^{ab}$ and $\textit{H}$$_{2}$$^{ab}$, respectively, and a tiny kink at the field $\textit{H}$$_{3}$$^{ab}$, indicating three transitions emerging, which will be discussed in details below. With decreasing temperature, the critical field $\textit{H}$$_{2}$$^{ab}$ shifts to a higher value, while the $\textit{H}$$_{1}$$^{ab}$ remains almost unchanged. To get more information on the abnormal jumps, we present the $\textit{M}$$_{ab}$ data measured at $\textit{T}$ = 2 K in the upper panel of Fig. 3(c). With increasing field, the $\textit{M}$$_{ab}$ increases linearly first, which is consistent with the expectation of AFM ground state. Then, it changes sharply at $\mu_{0}$$\textit{H}$$_{1}$$^{ab}$ = 0.53 T /0.44 T, when increasing/decreasing magnetic field, respectively, $\textit{i.e.}$, a hysteresis emerging at this field, corresponding to the first MM transition (MM-I). After the transition, the $\textit{M}$$_{ab}$($\textit{H}$) displays a linear dependence. With the field increasing further, the $\textit{M}$$_{ab}$($\textit{H}$) increases again sharply at $\mu_{0}$$\textit{H}$$_{2}$$^{ab}$ = 0.95 T without hysteresis, corresponding to a second MM transition (MM-II), and then turns to display a linear field dependence again. Finally, it saturates to 7.05 $\mu$$_{B}$ at $\mu_{0}$$\textit{H}$$_{3}$$^{ab}$ = 2.53 T. The critical fields are shown clearer from the derivative plot of $\textit{M}$$_{ab}$($\textit{H}$) as shown in the lower panel of Fig. 3(c). A hysteresis emerges at the MM-I transition, $\textit{i.e.}$, $M_{ab}$ values measured in the increasing and decreasing field process are not the same, indicating that this transition is the first order. Similar behavior has been reported in the polycrystalline EuAg$_{4}$As$_{2}$ samples previously, which was explained as a MM transition of the Eu$^{2+}$ moments. However, only one MM transition was observed in this work \cite{B.Gerke}, different from our results. In addition, we note that the $M_{ab}(H)$ curves for $\textit{T}$ $>$ 2 K still have a small slope in the high field region, which implies a FM-like alignment state rather than a FM ordered state. For comparison, we also measured the isothermal magnetization when the magnetic field is applied in the $\textit{c}$ axis, as shown in Fig. 3(e), but no MM transition was observed up to 7 T. At $\textit{T}$ = 2 K, the $\textit{M}$$_{c}$($\textit{H}$) increases slowly with increasing field compared with $\textit{M}$$_{ab}$($\textit{H}$), and displays two kinks around $\mu_{0}$$\textit{H}$$_{1}$$^{c}$ = 2.06 T, and $\mu_{0}$$\textit{H}$$_{2}$$^{c}$ = 4.14 T, respectively. A small slop is also observed in the high field region for $M_{c}$($\textit{H}$), which doesn't saturate until the highest measured field. At $\textit{T}$$_{N1}$ = 10 K, the kink in $M_{c}$($\textit{H}$) disappears, which may be related to the difference between AFM-I to AFM-II.

\begin{figure}
  \includegraphics[width=8cm]{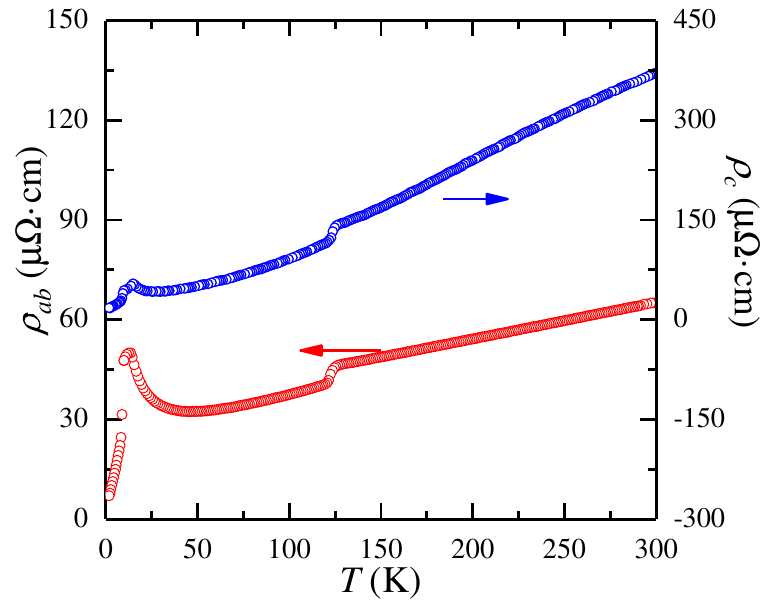}\\
  \caption{(Color online) Temperature dependence of the in-plane (left) and out-plane (right) resistivity of EuAg$_{4}$As$_{2}$ single crystal.}
\end{figure}

\begin{figure}
  \includegraphics[width=8cm]{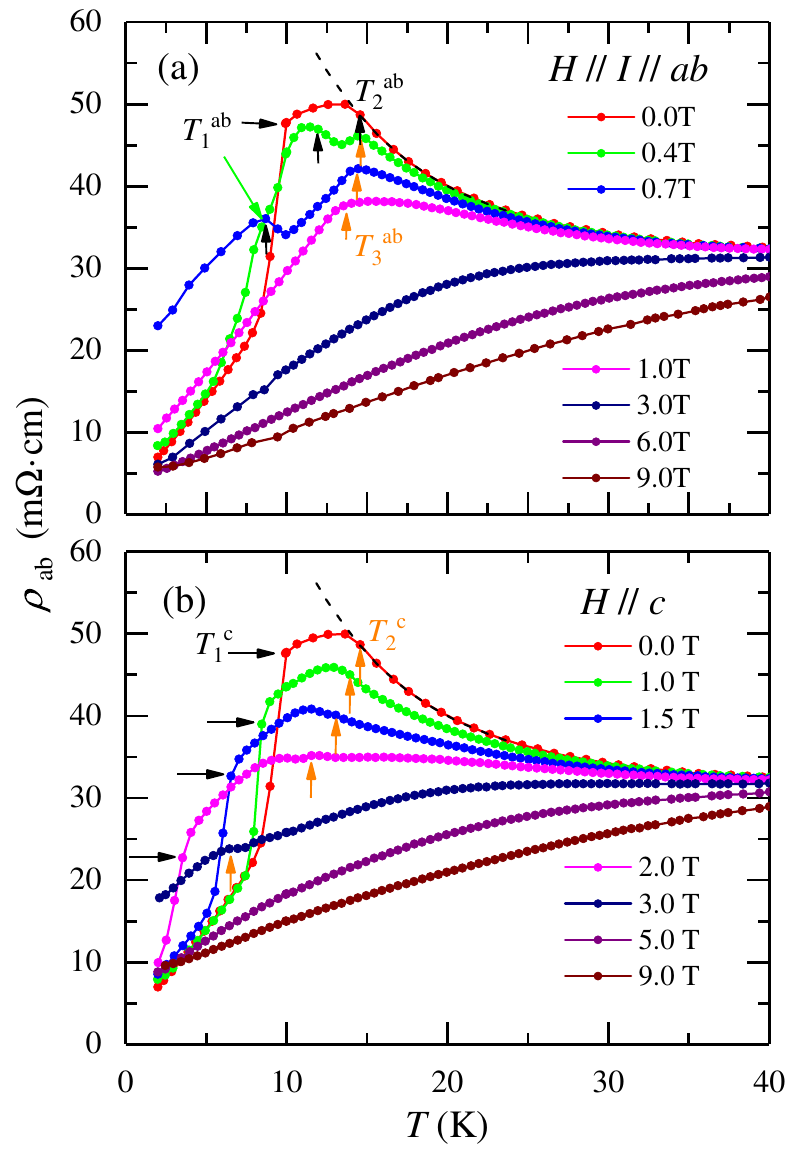}\\
  \caption{(Color online) Temperature dependence of in-plane resistivity $\rho$$_{ab}$ of EuAg$_{4}$As$_{2}$ single crystal under several selected fields for $\textit{H}$ $\parallel$ $\textit{ab}$ (a) and $\textit{H}$ $\parallel$ $\textit{c}$ (b), below $\textit{T}$ = 40 K. The dashed lines are guides to the eyes.}
\end{figure}

\begin{figure*}
  \includegraphics[width=14cm]{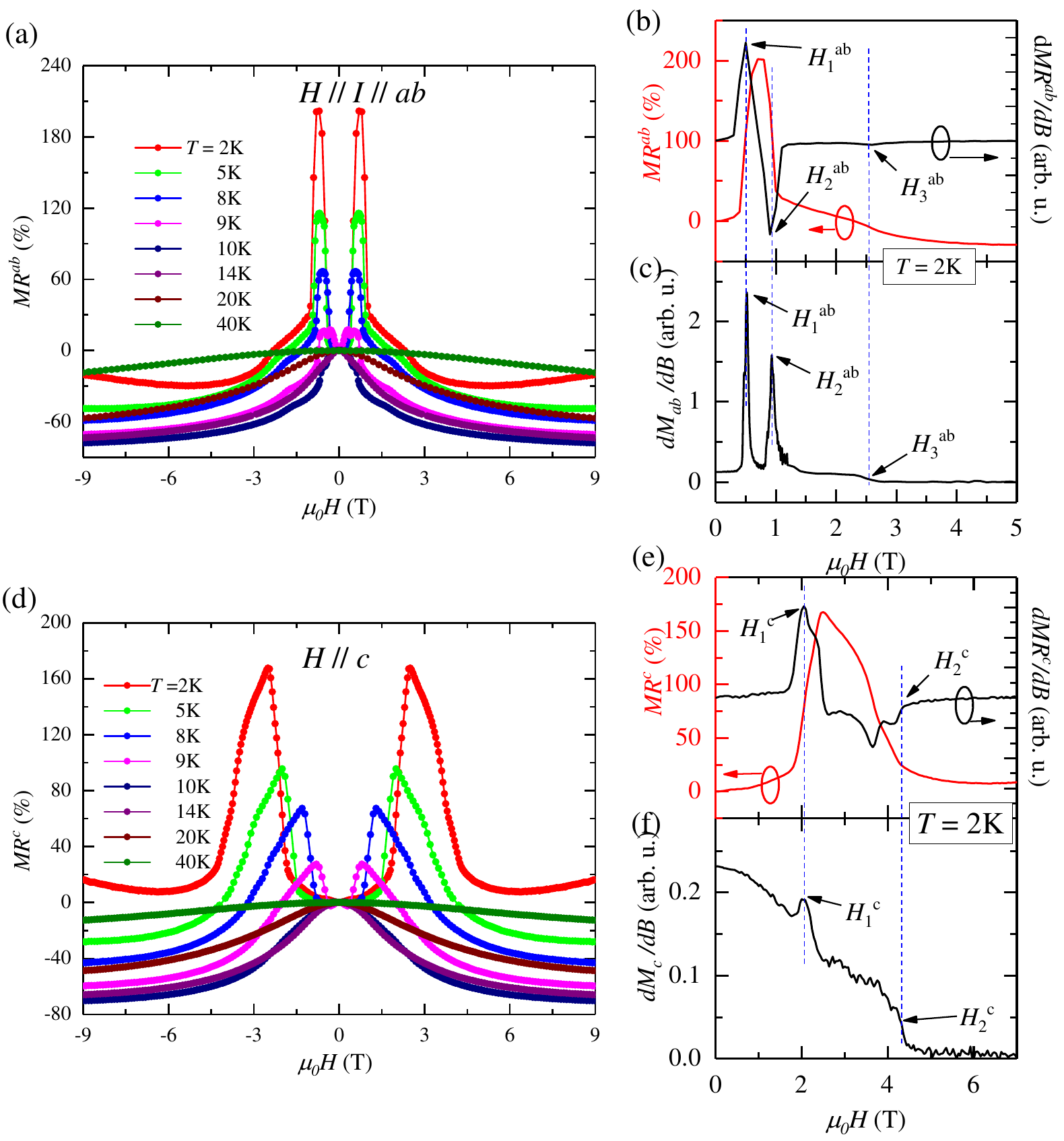}\\
  \caption{(Color online) Field dependence of the MR at several temperatures for $\textit{H}$ $\parallel$ $\textit{ab}$ (a) and $\textit{H}$ $\parallel$ $\textit{c}$ (d). The applied current is parallel to the field in the case of $\textit{H}$ $\parallel$ $\textit{ab}$. (b) The field dependence of MR$^{ab}$ (left) and its first derivative (right) at $\textit{T}$ = 2 K. (c) The first derivative of the $\textit{M}$$_{ab}$ as a function of field at $\textit{T}$ = 2 K. (e) The field dependence of MR$^{c}$ (left) and its first derivative (right) at $\textit{T}$ = 2 K. (f) The first derivative of the $\textit{M}$$_{c}$ as a function of field at $\textit{T}$ = 2 K. The dashed lines are guides to the eyes.}
\end{figure*}

Next, we discuss the two MM transitions observed in $\textit{M}$$_{ab}$($\textit{H}$) curves. According to the neutron diffraction result reported in Ref. \cite{B.Shen2}, below $\textit{T}$$_{N1}$,  EuAg$_{4}$As$_{2}$ is an incommesurate antiferromagnet (AFM-I), $\textit{i.e.}$, the spins of Eu$^{2+}$ rotate around the $\textit{c}$ axis with a helical arrangement, and around the $\textit{b}$ axis with a cycloidal arrangement, having a small propagation vector of $\vec{K}$$_{m}$ = (0,-0.1,0.12). Such a complex non-collinear magnetic structure was considered to be resulted from the RKKY interaction in EuAg$_{4}$As$_{2}$ \cite{B.Shen2}. As discussed above, with increasing field applied in the $\textit{ab}$ plane, the $\textit{M}$$_{ab}$ undergoes two steep jumps at $\textit{H}$$_{1}$$^{ab}$ and $\textit{H}$$_{2}$$^{ab}$, respectively, the first jump exhibiting a notable hysteresis, while the second one without a hysteresis, then saturates finally to 7.05 $\mu$$_{B}$ at $\textit{H}$$_{3}$$^{ab}$. However, no MM transition was observed as $\textit{H}$ was applied in the $\textit{c}$ axis. These results indicated that the spin-alignment of EuAg$_{4}$As$_{2}$ for $\textit{H}$ $\parallel$ $\textit{ab}$ may transit into two metastable states (MSS-I and MSS-II) successively with increasing field. Although the exact arrangement of spins in these metastable states can't be described only by our static magnetization measurements, and is needed to be determined by the neutron diffraction experiments in the future, the corresponding MM transitions are different from that due to the spin-flop (in which the spins are driven perpendicular to the applied magnetic field), which were usually observed in an uniaxial antiferromagnet with low anisotropy, such as in CaCo$_{2}$As$_{2}$ \cite{B.Cheng}. In EuAg$_{4}$As$_{2}$, the spin-alignment may have four states, AFM-I (discussed above), MSS-I ( $\textit{H}$$_{1}$$^{ab}$ $<$ H $<$ $\textit{H}$$_{2}$$^{ab}$), MSS-II ( $\textit{H}$$_{2}$$^{ab}$ $<$ H $<$ $\textit{H}$$_{3}$$^{ab}$), and FM-like alignment (H $>$ $\textit{H}$$_{3}$$^{ab}$) under applied magnetic field in the $\textit{ab}$ plane. Finally, the balance between Zeeman energy, magnetic coupling energy, and magneto-crystalline anisotropy energy lead to that the moments are rotated to the direction of $\textit{H}$ (in the $\textit{ab}$ plane). Similar phenomena can also be found in the other rare-earth intermetallic compounds \cite{Anupam, J.Tong, S.Jiang, F.Weber}.

Figure 4 shows the electrical resistivity in the $\textit{ab}$-plane, $\rho_{ab}$($\textit{T}$), and along $\textit{c}$ axis, $\rho_{c}$($\textit{T}$), as a function of temperature for a EuAg$_{4}$As$_{2}$ crystal. The $\rho_{ab}$ and $\rho_c$ at room temperature (300 K) are of 65 $\mu$$\Omega$ $cm$ and 370 $\mu$$\Omega$ $cm$, respectively, thus the resistivity anisotropy $\rho_{c}$/$\rho_{ab}$ = 5.7, which is not so large for this layered compound. With decreasing temperature from 300 K, the $\rho_{ab}$ deceases monotonically at first, then drops sharply near $T_s$ = 120 K due to a structural transition \cite{B.Shen2}, and goes through a hump around 15 K, which can be ascribed to the magnetic transitions. The $\rho_c(T)$ exhibits a similar behavior.

Then, we discuss the magnetic responses of $\rho_{ab}(T)$ measured at low temperatures ($\leq$ 40 K) with applied $H$ $\parallel$ $ab$ plane and $H$ $\parallel$ $c$ axis, respectively. Under zero field, the $\rho_{ab}$($\textit{T}$) exhibits a hump feature starting at $\textit{T}$$_{N2}$ = 15 K, and a rapid drop around $\textit{T}$$_{N1}$ = 10 K, which is consistent with the magnetic transitions. For both $\textit{H}$ $\parallel$ $\textit{ab}$ and $\textit{H}$ $\parallel$ $\textit{c}$, the two transitions shifts to lower temperatures with increasing field, and the resistivity hump is suppressed, resulting in a large MR. Meanwhile, an additional transition is observed at $\textit{T}$$_{3}$$^{ab}$ in the case of $\textit{H}$ $\parallel$ $\textit{ab}$, characterized by a peak (or kink) in the $\rho$$_{ab}$($\textit{T}$) curves, consistent with the $\chi$$_{ab}$(\textit{T}) data as discussed above. In the highest measured field of 9 T, the $\rho_{ab}$($\textit{T}$) deceases monotonically with decreasing temperature, and no phase transition is observed for both $\textit{H}$ $\parallel$ $\textit{ab}$ and $\textit{H}$ $\parallel$ $\textit{c}$.

Figure 6(a) shows the field dependence of MR for $\textit{H}$ $\parallel$ $\textit{ab}$ measured at several temperatures. In order to eliminate the hysteresis effect in the first MM transition region, all the data are collected in a field increasing process. At $\textit{T}$ = 2 K, the MR$^{ab}$, defined as $\frac{\rho_{ab}(\textit{H}, \textit{T})-\rho_{ab}(0, \textit{T})}{\rho_{ab}(0, \textit{T})}$, increases slowly with increasing field at first, then displays a quick rise around $\mu_{0}$$\textit{H}$$_{1}$$^{ab}$ = 0.5 T, and reaches a maximum value of 202$\%$ at 0.7 T, then decreases rapidly until $\mu_{0}$$\textit{H}$ = 1 T, exhibiting a peak-like feature. With increasing field further, the MR$^{ab}$ decreases gradually to negative values, reaches a minimum, then increases a little, consistent with the behavior of the FM-like state. The critical fields at $\textit{T}$ = 2 K are clearly shown in the first derivative of MR$^{ab}$, and are also consistent with that of d\textit{M}$_{ab}$/d\textit{H} [see Fig. 6(b) and 6(c)], indicating that the spin-alignment of MSS-I has the strongest scatter to electrons, resulting in a large positive MR. With increasing temperature, the positive MR$^{ab}$ at low fields is significantly suppressed, and disappears for $\textit{T}$ $\geq$ 10 K. Instead, a large negative MR$^{ab}$ emerges for $\textit{T}$ $>$ 10 K in the whole measuring field range, which is probably contributed to the reduction of spin disorder scattering. At $\textit{T}$ = 10 K, the MR$^{ab}$ can even reach -78$\%$ at 9 T. With increasing temperature further, the magnitude of the negative MR$^{ab}$ decreases, exhibiting a maximum at $\textit{T}$$_{N1}$. This behavior is similar to that observed in the well-known perovskites-based manganites (CMR systems) \cite{R.Helmolt}. It is interesting that the negative MR$^{ab}$ is also observed at higher temperatures far above $\textit{T}$$_{N2}$ in EuAg$_{4}$As$_{2}$, such as, the MR$^{ab}$ reaches as large as -21$\%$ at 9 T for $\textit{T}$ = 40 K, as shown in Fig. 6(a), which usually occurs in the ferromagnetically ordered state. We also note that there is a sign crossover around 65 K, beyond which MR$^{ab}$ is a negligibly small, but positive value (not shown here). So we suggest that the large negative MR above $\textit{T}$$_{N2}$ may origin from the precursor effect of Eu$^{2+}$ 4f$^{7}$ moment long-range magnetic ordering, as discussed in Eu$_{2}$CuSi$_{3}$ and Eu$_{3}$Ni$_{4}$Ga$_{4}$ \cite{S.Majumdar, Anupam}.

\begin{figure}
  \includegraphics[width=8cm]{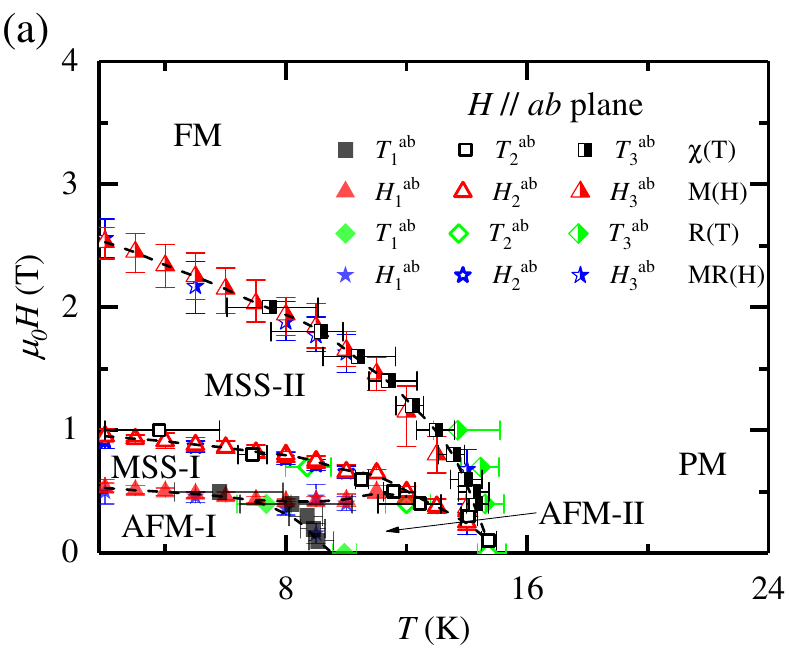}\\
  \includegraphics[width=8cm]{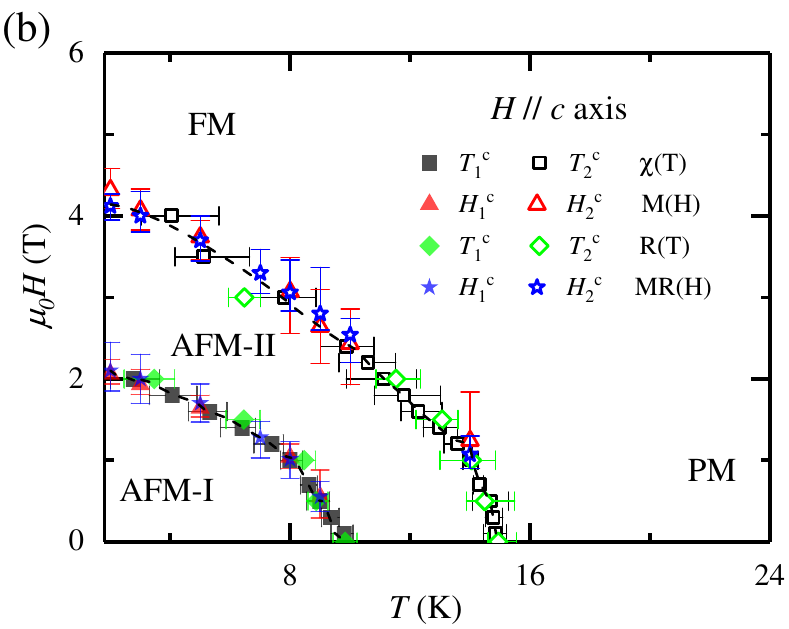}\\
  \caption{Magnetic phase diagram of EuAg$_{4}$As$_{2}$ for $\textit{H}$ $\parallel$ $\textit{ab}$ plane (a) and $\textit{H}$ $\parallel$ $\textit{c}$ axis (b). The symbols are extracted from $\textit{M}$-$\textit{T}$ (black), $\textit{M}$-$\textit{H}$ (red), $\textit{R}$-$\textit{T}$ (green) and MR-$\textit{H}$ (blue) curves, respectively. The dashed lines are guides to the eyes.}
\end{figure}

As shown in Fig. 6(d), the field dependence of MR$^{c}$ ($\textit{H}$ $\parallel$ $\textit{c}$) is quite similar to the behavior in MR$^{ab}$($\textit{H}$). At $\textit{T}$ = 2 K, with increasing field, the MR$^{c}$ increases gradually at first, then increases sharply at $\mu_{0}$$\textit{H}$$_{1}$$^{c}$ = 2.1 T, goes through a maximum (over 160$\%$ at 2 K and 2.5 T), and decreases steeply until $\mu_{0}$$\textit{H}$$_{2}$$^{c}$ = 4.1 T, reaches a minimum, finally increases a little until the highest measuring field (9 T). The MR$^{c}$ behavior is consistent with that of \textit{M}$_{c}$(\textit{H}), as shown in Fig. 6(f). Compared with d\textit{M}$_{c}$/d\textit{B}, several additional peaks were observed in the first derivative of MR$^{c}$, which may be ascribed to the movement of magnetic domain walls with the external field. These results imply that when the magnetic field is applied along the $\textit{c}$ axis, the spin alignment in AFM-II has the strongest scatter to electrons flowing in the $\textit{ab}$ plane (corresponding to $\rho$$_{ab}$), resulting in a large positive MR. With increasing temperature, the $\textit{H}$$_{1}$$^{c}$ and $\textit{H}$$_{2}$$^{c}$, and the positive MR decrease. At $\textit{T}$ = 10 K, the peak of positive MR disappears, and the largest negative MR (up to -70$\%$ at 9 T) emerges. Up to 40K, far above $\textit{T}$$_{N2}$, a large negative MR remains (-20$\%$, 9T). Another, except for 2 K, the MR decreases monotonically with increasing magnetic field after the positive peak at all temperatures, which maybe related the spin dynamics.

As discussed above, such complicated behaviors exhibiting in $\rho_{ab}$($\textit{T}$) and $\rho_{c}$($\textit{T}$) in EuAg$_{4}$As$_{2}$ crystal  are related to the magnetic ground states at different magnetic fields (\textit{H}) and various temperatures (\textit{T}). In order to clarify the relationship between the transport and magnetic order, we construct the \textit{H}(\textit{T}) phase diagram based on the resistivity and magnetization data measured at various (\textit{H}, \textit{T}), as shown in Fig. 7(a) ($\textit{H}$ $\parallel$ $\textit{ab}$) and Fig. 7(b) ($\textit{H}$ $\parallel$ $\textit{c}$), respectively. For $\textit{H}$ $\parallel$ $\textit{ab}$ plane, the phase diagram can be divided into six regions, AFM-I, AFM-II, PM, FM-like alignment region, and two metastable states (MSS-I and MSS-II). At $\textit{T}$ = 2 K, the compound undergoes two MM transitions from AFM-I with increasing field and enters finally the FM-like alignment region, with the critical field of 0.5 T, 0.95 T and 2.5 T, respectively. It should be noted that the boundary between FM-like alignment region and PM state can't be precisely determined. For $\textit{H}$ $\parallel$ $\textit{c}$ axis, the phase diagram can be divided into four regions, AFM-I, AFM-II, PM and FM-like alignment region, two phase boundaries are clearly distinguished. At $\textit{T}$ = 2 K, EuAg$_{4}$As$_{2}$ crystal undergoes magnetic phase transitions from the AFM-I phase to the intermediate AFM-II phase, then to the FM-like alignment region with increasing magnetic field, with the critical field of 2.1 T, and 4.2 T, respectively. Another, as shown in Fig. 7(a) and Fig. 7(b), the phase boundaries deduced from the data of magnetization and resistivity measurements, respectively, are well consistent with each other, indicating that the complicated magnetotransport properties are related to the magnetic orders, especially the large positive MR occurring in a narrow magnetic range provides a chance for application on magnetic sensors.

In summary, we studied systematically the magnetic and transport properties of EuAg$_{4}$As$_{2}$ single crystals by using magnetization and resistivity measurements. Under zero field, it was confirmed that two magnetic transitions occur at $\textit{T}$$_{N1}$ = 10 K and $\textit{T}$$_{N2}$ = 15 K, respectively, with the magnetic moments almost lying in the $\textit{ab}$ plane. With increasing field, the two magnetic transitions are driven noticeably to lower temperatures, indicating that they are tunable ground states. At $\textit{T}$ = 2 K, two successive MM transitions were observed at 0.5 T and 0.95 T, respectively, when applying magnetic field in the $\textit{ab}$ plane. On the other hand, no MM transition was detected below 7 T for $\textit{H}$ $\parallel$ $\textit{c}$. For both $\textit{H}$ $\parallel$ $\textit{ab}$ and $\textit{H}$ $\parallel$ $\textit{c}$, an anomalous field dependence of MR was observed, which shows a positive, unexpected large value at low fields below 10 K, and a large negative value at high fields/intermediate temperatures, indicating a rather disordered spin-alignment state in the intermediate phases. Such anomalous field dependence of MR is rather rare, and may have potential application in the future magnetic sensors. Interestingly, large negative MR is seen even at 40 K, which is far above the magnetic transition temperature. According to these results, we established the phase diagrams of EuAg$_{4}$As$_{2}$ for both $\textit{H}$ $\parallel$ $\textit{ab}$ and $\textit{H}$ $\parallel$ $\textit{c}$.

\section{ACKNOWLEDGEMENTS}
This research is supported by the Ministry of Science and Technology of China under Grants No. 2016YFA0300402 and No. 2015CB921004 and the National Natural Science Foundation of China (NSFC) (No. 11974095, 11374261), the Zhejiang Natural Science Foundation (No. LY16A040012) and the Fundamental Research Funds for the Central Universities.

\bibliography{EuAg4As2}

\begin{thebibliography}{18}%
\makeatletter
\providecommand \@ifxundefined [1]{%
 \@ifx{#1\undefined}
}%
\providecommand \@ifnum [1]{%
 \ifnum #1\expandafter \@firstoftwo
 \else \expandafter \@secondoftwo
 \fi
}%
\providecommand \@ifx [1]{%
 \ifx #1\expandafter \@firstoftwo
 \else \expandafter \@secondoftwo
 \fi
}%
\providecommand \natexlab [1]{#1}%
\providecommand \enquote  [1]{``#1''}%
\providecommand \bibnamefont  [1]{#1}%
\providecommand \bibfnamefont [1]{#1}%
\providecommand \citenamefont [1]{#1}%
\providecommand \href@noop [0]{\@secondoftwo}%
\providecommand \href [0]{\begingroup \@sanitize@url \@href}%
\providecommand \@href[1]{\@@startlink{#1}\@@href}%
\providecommand \@@href[1]{\endgroup#1\@@endlink}%
\providecommand \@sanitize@url [0]{\catcode `\\12\catcode `\$12\catcode
  `\&12\catcode `\#12\catcode `\^12\catcode `\_12\catcode `\%12\relax}%
\providecommand \@@startlink[1]{}%
\providecommand \@@endlink[0]{}%
\providecommand \url  [0]{\begingroup\@sanitize@url \@url }%
\providecommand \@url [1]{\endgroup\@href {#1}{\urlprefix }}%
\providecommand \urlprefix  [0]{URL }%
\providecommand \Eprint [0]{\href }%
\providecommand \doibase [0]{http://dx.doi.org/}%
\providecommand \selectlanguage [0]{\@gobble}%
\providecommand \bibinfo  [0]{\@secondoftwo}%
\providecommand \bibfield  [0]{\@secondoftwo}%
\providecommand \translation [1]{[#1]}%
\providecommand \BibitemOpen [0]{}%
\providecommand \bibitemStop [0]{}%
\providecommand \bibitemNoStop [0]{.\EOS\space}%
\providecommand \EOS [0]{\spacefactor3000\relax}%
\providecommand \BibitemShut  [1]{\csname bibitem#1\endcsname}%
\let\auto@bib@innerbib\@empty
\bibitem [{\citenamefont {Mitsuda}\ \emph {et~al.}(1997)\citenamefont
  {Mitsuda}, \citenamefont {Wada}, \citenamefont {Shiga}, \citenamefont
  {Aruga~Katori},\ and\ \citenamefont {Goto}}]{A.Mitsuda}%
  \BibitemOpen
  \bibfield  {author} {\bibinfo {author} {\bibfnamefont {A.}~\bibnamefont
  {Mitsuda}}, \bibinfo {author} {\bibfnamefont {H.}~\bibnamefont {Wada}},
  \bibinfo {author} {\bibfnamefont {M.}~\bibnamefont {Shiga}}, \bibinfo
  {author} {\bibfnamefont {H.}~\bibnamefont {Aruga~Katori}}, \ and\ \bibinfo
  {author} {\bibfnamefont {T.}~\bibnamefont {Goto}},\ }\href {\doibase
  10.1103/PhysRevB.55.12474} {\bibfield  {journal} {\bibinfo  {journal} {Phys.
  Rev. B}\ }\textbf {\bibinfo {volume} {55}},\ \bibinfo {pages} {12474}
  (\bibinfo {year} {1997})}\BibitemShut {NoStop}%
\bibitem [{\citenamefont {Feng}\ \emph {et~al.}(2010)\citenamefont {Feng},
  \citenamefont {Ren}, \citenamefont {Xu}, \citenamefont {Jiang}, \citenamefont
  {Xu}, \citenamefont {Cao}, \citenamefont {Nowik}, \citenamefont {Felner},
  \citenamefont {Matsubayashi},\ and\ \citenamefont {Uwatoko}}]{C.Feng}%
  \BibitemOpen
  \bibfield  {author} {\bibinfo {author} {\bibfnamefont {C.}~\bibnamefont
  {Feng}}, \bibinfo {author} {\bibfnamefont {Z.}~\bibnamefont {Ren}}, \bibinfo
  {author} {\bibfnamefont {S.}~\bibnamefont {Xu}}, \bibinfo {author}
  {\bibfnamefont {S.}~\bibnamefont {Jiang}}, \bibinfo {author} {\bibfnamefont
  {Z.}~\bibnamefont {Xu}}, \bibinfo {author} {\bibfnamefont {G.}~\bibnamefont
  {Cao}}, \bibinfo {author} {\bibfnamefont {I.}~\bibnamefont {Nowik}}, \bibinfo
  {author} {\bibfnamefont {I.}~\bibnamefont {Felner}}, \bibinfo {author}
  {\bibfnamefont {K.}~\bibnamefont {Matsubayashi}}, \ and\ \bibinfo {author}
  {\bibfnamefont {Y.}~\bibnamefont {Uwatoko}},\ }\href {\doibase
  10.1103/PhysRevB.82.094426} {\bibfield  {journal} {\bibinfo  {journal} {Phys.
  Rev. B}\ }\textbf {\bibinfo {volume} {82}},\ \bibinfo {pages} {094426}
  (\bibinfo {year} {2010})}\BibitemShut {NoStop}%
\bibitem [{\citenamefont {Masuda}\ \emph {et~al.}(2016)\citenamefont {Masuda},
  \citenamefont {Sakai}, \citenamefont {Tokunaga}, \citenamefont {Yamasaki},
  \citenamefont {Miyake}, \citenamefont {Shiogai}, \citenamefont {Nakamura},
  \citenamefont {Awaji}, \citenamefont {Tsukazaki}, \citenamefont {Nakao},
  \citenamefont {Murakami}, \citenamefont {Arima}, \citenamefont {Tokura},\
  and\ \citenamefont {Ishiwata}}]{H.Masuda}%
  \BibitemOpen
  \bibfield  {author} {\bibinfo {author} {\bibfnamefont {H.}~\bibnamefont
  {Masuda}}, \bibinfo {author} {\bibfnamefont {H.}~\bibnamefont {Sakai}},
  \bibinfo {author} {\bibfnamefont {M.}~\bibnamefont {Tokunaga}}, \bibinfo
  {author} {\bibfnamefont {Y.}~\bibnamefont {Yamasaki}}, \bibinfo {author}
  {\bibfnamefont {A.}~\bibnamefont {Miyake}}, \bibinfo {author} {\bibfnamefont
  {J.}~\bibnamefont {Shiogai}}, \bibinfo {author} {\bibfnamefont
  {S.}~\bibnamefont {Nakamura}}, \bibinfo {author} {\bibfnamefont
  {S.}~\bibnamefont {Awaji}}, \bibinfo {author} {\bibfnamefont
  {A.}~\bibnamefont {Tsukazaki}}, \bibinfo {author} {\bibfnamefont
  {H.}~\bibnamefont {Nakao}}, \bibinfo {author} {\bibfnamefont
  {Y.}~\bibnamefont {Murakami}}, \bibinfo {author} {\bibfnamefont {T.-h.}\
  \bibnamefont {Arima}}, \bibinfo {author} {\bibfnamefont {Y.}~\bibnamefont
  {Tokura}}, \ and\ \bibinfo {author} {\bibfnamefont {S.}~\bibnamefont
  {Ishiwata}},\ }\href {\doibase 10.1126/sciadv.1501117} {\bibfield  {journal}
  {\bibinfo  {journal} {Sci. Adv.}\ }\textbf {\bibinfo {volume} {2}},\ \bibinfo
  {pages} {e1501117} (\bibinfo {year} {2016})}\BibitemShut {NoStop}%
\bibitem [{\citenamefont {Majumdar}\ \emph {et~al.}(1999)\citenamefont
  {Majumdar}, \citenamefont {Mallik}, \citenamefont {Sampathkumaran},
  \citenamefont {Rupprecht},\ and\ \citenamefont {Wortmann}}]{S.Majumdar}%
  \BibitemOpen
  \bibfield  {author} {\bibinfo {author} {\bibfnamefont {S.}~\bibnamefont
  {Majumdar}}, \bibinfo {author} {\bibfnamefont {R.}~\bibnamefont {Mallik}},
  \bibinfo {author} {\bibfnamefont {E.~V.}\ \bibnamefont {Sampathkumaran}},
  \bibinfo {author} {\bibfnamefont {K.}~\bibnamefont {Rupprecht}}, \ and\
  \bibinfo {author} {\bibfnamefont {G.}~\bibnamefont {Wortmann}},\ }\href
  {\doibase 10.1103/PhysRevB.60.6770} {\bibfield  {journal} {\bibinfo
  {journal} {Phys. Rev. B}\ }\textbf {\bibinfo {volume} {60}},\ \bibinfo
  {pages} {6770} (\bibinfo {year} {1999})}\BibitemShut {NoStop}%
\bibitem [{\citenamefont {Yi}\ \emph {et~al.}(2017)\citenamefont {Yi},
  \citenamefont {Yang}, \citenamefont {Yang}, \citenamefont {Wang},
  \citenamefont {Matsushita}, \citenamefont {Miao}, \citenamefont {Jiao},
  \citenamefont {Cheng}, \citenamefont {Li}, \citenamefont {Yamaura},
  \citenamefont {Shi},\ and\ \citenamefont {Luo}}]{C.Yi}%
  \BibitemOpen
  \bibfield  {author} {\bibinfo {author} {\bibfnamefont {C.}~\bibnamefont
  {Yi}}, \bibinfo {author} {\bibfnamefont {S.}~\bibnamefont {Yang}}, \bibinfo
  {author} {\bibfnamefont {M.}~\bibnamefont {Yang}}, \bibinfo {author}
  {\bibfnamefont {L.}~\bibnamefont {Wang}}, \bibinfo {author} {\bibfnamefont
  {Y.}~\bibnamefont {Matsushita}}, \bibinfo {author} {\bibfnamefont
  {S.}~\bibnamefont {Miao}}, \bibinfo {author} {\bibfnamefont {Y.}~\bibnamefont
  {Jiao}}, \bibinfo {author} {\bibfnamefont {J.}~\bibnamefont {Cheng}},
  \bibinfo {author} {\bibfnamefont {Y.}~\bibnamefont {Li}}, \bibinfo {author}
  {\bibfnamefont {K.}~\bibnamefont {Yamaura}}, \bibinfo {author} {\bibfnamefont
  {Y.}~\bibnamefont {Shi}}, \ and\ \bibinfo {author} {\bibfnamefont
  {J.}~\bibnamefont {Luo}},\ }\href {\doibase 10.1103/PhysRevB.96.205103}
  {\bibfield  {journal} {\bibinfo  {journal} {Phys. Rev. B}\ }\textbf {\bibinfo
  {volume} {96}},\ \bibinfo {pages} {205103} (\bibinfo {year}
  {2017})}\BibitemShut {NoStop}%
\bibitem [{\citenamefont {Gignoux}\ and\ \citenamefont
  {Schmitt}(1991)}]{D.Gignoux}%
  \BibitemOpen
  \bibfield  {author} {\bibinfo {author} {\bibfnamefont {D.}~\bibnamefont
  {Gignoux}}\ and\ \bibinfo {author} {\bibfnamefont {D.}~\bibnamefont
  {Schmitt}},\ }\href {\doibase 10.1016/0304-8853(91)90815-R} {\bibfield
  {journal} {\bibinfo  {journal} {J. Magn. Magn. Mater.}\ }\textbf {\bibinfo
  {volume} {100}},\ \bibinfo {pages} {99} (\bibinfo {year} {1991})}\BibitemShut
  {NoStop}%
\bibitem [{\citenamefont {Stoyko}\ \emph {et~al.}(2012)\citenamefont {Stoyko},
  \citenamefont {Khatun}, \citenamefont {Mullen},\ and\ \citenamefont
  {Mar}}]{S.Stoyko}%
  \BibitemOpen
  \bibfield  {author} {\bibinfo {author} {\bibfnamefont {S.~S.}\ \bibnamefont
  {Stoyko}}, \bibinfo {author} {\bibfnamefont {M.}~\bibnamefont {Khatun}},
  \bibinfo {author} {\bibfnamefont {C.~S.}\ \bibnamefont {Mullen}}, \ and\
  \bibinfo {author} {\bibfnamefont {A.}~\bibnamefont {Mar}},\ }\href {\doibase
  10.1016/j.jssc.2012.04.042} {\bibfield  {journal} {\bibinfo  {journal} {J.
  Solid State Chem.}\ }\textbf {\bibinfo {volume} {192}},\ \bibinfo {pages}
  {325} (\bibinfo {year} {2012})}\BibitemShut {NoStop}%
\bibitem [{\citenamefont {Shen}\ \emph {et~al.}(2018)\citenamefont {Shen},
  \citenamefont {Emmanouilidou}, \citenamefont {Deng}, \citenamefont
  {McCollam}, \citenamefont {Xing}, \citenamefont {Kotliar}, \citenamefont
  {Coldea},\ and\ \citenamefont {Ni}}]{B.Shen}%
  \BibitemOpen
  \bibfield  {author} {\bibinfo {author} {\bibfnamefont {B.}~\bibnamefont
  {Shen}}, \bibinfo {author} {\bibfnamefont {E.}~\bibnamefont {Emmanouilidou}},
  \bibinfo {author} {\bibfnamefont {X.}~\bibnamefont {Deng}}, \bibinfo {author}
  {\bibfnamefont {A.}~\bibnamefont {McCollam}}, \bibinfo {author}
  {\bibfnamefont {J.}~\bibnamefont {Xing}}, \bibinfo {author} {\bibfnamefont
  {G.}~\bibnamefont {Kotliar}}, \bibinfo {author} {\bibfnamefont {A.~I.}\
  \bibnamefont {Coldea}}, \ and\ \bibinfo {author} {\bibfnamefont
  {N.}~\bibnamefont {Ni}},\ }\href {\doibase 10.1103/PhysRevB.98.235130}
  {\bibfield  {journal} {\bibinfo  {journal} {Phys. Rev. B}\ }\textbf {\bibinfo
  {volume} {98}},\ \bibinfo {pages} {235130} (\bibinfo {year}
  {2018})}\BibitemShut {NoStop}%
\bibitem [{\citenamefont {Gerke}\ \emph {et~al.}(2013)\citenamefont {Gerke},
  \citenamefont {Schwickert}, \citenamefont {Stoyko}, \citenamefont {Khatun},
  \citenamefont {Mar},\ and\ \citenamefont {Poettgen}}]{B.Gerke}%
  \BibitemOpen
  \bibfield  {author} {\bibinfo {author} {\bibfnamefont {B.}~\bibnamefont
  {Gerke}}, \bibinfo {author} {\bibfnamefont {C.}~\bibnamefont {Schwickert}},
  \bibinfo {author} {\bibfnamefont {S.~S.}\ \bibnamefont {Stoyko}}, \bibinfo
  {author} {\bibfnamefont {M.}~\bibnamefont {Khatun}}, \bibinfo {author}
  {\bibfnamefont {A.}~\bibnamefont {Mar}}, \ and\ \bibinfo {author}
  {\bibfnamefont {R.}~\bibnamefont {Poettgen}},\ }\href {\doibase
  10.1016/j.solidstatesciences.2013.02.027} {\bibfield  {journal} {\bibinfo
  {journal} {Solid State Sci.}\ }\textbf {\bibinfo {volume} {20}},\ \bibinfo
  {pages} {65} (\bibinfo {year} {2013})}\BibitemShut {NoStop}%
\bibitem [{\citenamefont {Shen}\ \emph {et~al.}()\citenamefont {Shen},
  \citenamefont {Hu}, \citenamefont {Cao}, \citenamefont {Gui}, \citenamefont
  {Emmanouilidou}, \citenamefont {Xie},\ and\ \citenamefont {Ni}}]{B.Shen2}%
  \BibitemOpen
  \bibfield  {author} {\bibinfo {author} {\bibfnamefont {B.}~\bibnamefont
  {Shen}}, \bibinfo {author} {\bibfnamefont {C.}~\bibnamefont {Hu}}, \bibinfo
  {author} {\bibfnamefont {H.}~\bibnamefont {Cao}}, \bibinfo {author}
  {\bibfnamefont {X.}~\bibnamefont {Gui}}, \bibinfo {author} {\bibfnamefont
  {E.}~\bibnamefont {Emmanouilidou}}, \bibinfo {author} {\bibfnamefont
  {W.}~\bibnamefont {Xie}}, \ and\ \bibinfo {author} {\bibfnamefont
  {N.}~\bibnamefont {Ni}},\ }\href@noop {} {}\Eprint
  {http://arxiv.org/abs/1809.07317} {arXiv:1809.07317 [cond-mat.str-el]}
  \BibitemShut {NoStop}%
\bibitem [{\citenamefont {Ryan}\ \emph {et~al.}(2019)\citenamefont {Ryan},
  \citenamefont {Bud'ko}, \citenamefont {Hu},\ and\ \citenamefont
  {N.}}]{D.H.Ryan}%
  \BibitemOpen
  \bibfield  {author} {\bibinfo {author} {\bibfnamefont {D.~H.}\ \bibnamefont
  {Ryan}}, \bibinfo {author} {\bibfnamefont {S.~L.}\ \bibnamefont {Bud'ko}},
  \bibinfo {author} {\bibfnamefont {C.}~\bibnamefont {Hu}}, \ and\ \bibinfo
  {author} {\bibfnamefont {N.}~\bibnamefont {N.}},\ }\href@noop {} {\bibfield
  {journal} {\bibinfo  {journal} {AIP Adv.}\ }\textbf {\bibinfo {volume} {9}},\
  \bibinfo {pages} {125050} (\bibinfo {year} {2019})}\BibitemShut {NoStop}%
\bibitem [{\citenamefont {Bud'ko}\ \emph {et~al.}(2020)\citenamefont {Bud'ko},
  \citenamefont {Xiang}, \citenamefont {Hu}, \citenamefont {Shen},
  \citenamefont {Ni},\ and\ \citenamefont {Canfield}}]{L.Sergey}%
  \BibitemOpen
  \bibfield  {author} {\bibinfo {author} {\bibfnamefont {S.~L.}\ \bibnamefont
  {Bud'ko}}, \bibinfo {author} {\bibfnamefont {L.}~\bibnamefont {Xiang}},
  \bibinfo {author} {\bibfnamefont {C.}~\bibnamefont {Hu}}, \bibinfo {author}
  {\bibfnamefont {B.}~\bibnamefont {Shen}}, \bibinfo {author} {\bibfnamefont
  {N.}~\bibnamefont {Ni}}, \ and\ \bibinfo {author} {\bibfnamefont {P.~C.}\
  \bibnamefont {Canfield}},\ }\href@noop {} {\bibfield  {journal} {\bibinfo
  {journal} {Phys. Rev. B}\ }\textbf {\bibinfo {volume} {101}},\ \bibinfo
  {pages} {195112} (\bibinfo {year} {2020})}\BibitemShut {NoStop}%
\bibitem [{\citenamefont {Cheng}\ \emph {et~al.}(2012)\citenamefont {Cheng},
  \citenamefont {Hu}, \citenamefont {Yuan}, \citenamefont {Dong}, \citenamefont
  {Fang}, \citenamefont {Chen}, \citenamefont {Xu}, \citenamefont {Shi},
  \citenamefont {Zheng}, \citenamefont {Luo},\ and\ \citenamefont
  {Wang}}]{B.Cheng}%
  \BibitemOpen
  \bibfield  {author} {\bibinfo {author} {\bibfnamefont {B.}~\bibnamefont
  {Cheng}}, \bibinfo {author} {\bibfnamefont {B.~F.}\ \bibnamefont {Hu}},
  \bibinfo {author} {\bibfnamefont {R.~H.}\ \bibnamefont {Yuan}}, \bibinfo
  {author} {\bibfnamefont {T.}~\bibnamefont {Dong}}, \bibinfo {author}
  {\bibfnamefont {A.~F.}\ \bibnamefont {Fang}}, \bibinfo {author}
  {\bibfnamefont {Z.~G.}\ \bibnamefont {Chen}}, \bibinfo {author}
  {\bibfnamefont {G.}~\bibnamefont {Xu}}, \bibinfo {author} {\bibfnamefont
  {Y.~G.}\ \bibnamefont {Shi}}, \bibinfo {author} {\bibfnamefont
  {P.}~\bibnamefont {Zheng}}, \bibinfo {author} {\bibfnamefont {J.~L.}\
  \bibnamefont {Luo}}, \ and\ \bibinfo {author} {\bibfnamefont {N.~L.}\
  \bibnamefont {Wang}},\ }\href {\doibase 10.1103/PhysRevB.85.144426}
  {\bibfield  {journal} {\bibinfo  {journal} {Phys. Rev. B}\ }\textbf {\bibinfo
  {volume} {85}},\ \bibinfo {pages} {144426} (\bibinfo {year}
  {2012})}\BibitemShut {NoStop}%
\bibitem [{\citenamefont {Anupam}\ \emph {et~al.}(2012)\citenamefont {Anupam},
  \citenamefont {Geibel},\ and\ \citenamefont {Hossain}}]{Anupam}%
  \BibitemOpen
  \bibfield  {author} {\bibinfo {author} {\bibnamefont {Anupam}}, \bibinfo
  {author} {\bibfnamefont {C.}~\bibnamefont {Geibel}}, \ and\ \bibinfo {author}
  {\bibfnamefont {Z.}~\bibnamefont {Hossain}},\ }\href@noop {} {\bibfield
  {journal} {\bibinfo  {journal} {J. Phys.: Conden. Matter}\ }\textbf {\bibinfo
  {volume} {24}},\ \bibinfo {pages} {326002} (\bibinfo {year}
  {2012})}\BibitemShut {NoStop}%
\bibitem [{\citenamefont {Tong}\ \emph {et~al.}(2014)\citenamefont {Tong},
  \citenamefont {Parry}, \citenamefont {Tao}, \citenamefont {Cao},
  \citenamefont {Xu},\ and\ \citenamefont {Zeng}}]{J.Tong}%
  \BibitemOpen
  \bibfield  {author} {\bibinfo {author} {\bibfnamefont {J.}~\bibnamefont
  {Tong}}, \bibinfo {author} {\bibfnamefont {J.}~\bibnamefont {Parry}},
  \bibinfo {author} {\bibfnamefont {Q.}~\bibnamefont {Tao}}, \bibinfo {author}
  {\bibfnamefont {G.-H.}\ \bibnamefont {Cao}}, \bibinfo {author} {\bibfnamefont
  {Z.-A.}\ \bibnamefont {Xu}}, \ and\ \bibinfo {author} {\bibfnamefont
  {H.}~\bibnamefont {Zeng}},\ }\href {\doibase 10.1016/j.jallcom.2014.02.157}
  {\bibfield  {journal} {\bibinfo  {journal} {J. Alloys Compnd.}\ }\textbf
  {\bibinfo {volume} {602}},\ \bibinfo {pages} {26} (\bibinfo {year}
  {2014})}\BibitemShut {NoStop}%
\bibitem [{\citenamefont {Jiang}\ \emph {et~al.}(2009)\citenamefont {Jiang},
  \citenamefont {Luo}, \citenamefont {Ren}, \citenamefont {Zhu}, \citenamefont
  {Wang}, \citenamefont {Xu}, \citenamefont {Tao}, \citenamefont {Cao},\ and\
  \citenamefont {Xu}}]{S.Jiang}%
  \BibitemOpen
  \bibfield  {author} {\bibinfo {author} {\bibfnamefont {S.}~\bibnamefont
  {Jiang}}, \bibinfo {author} {\bibfnamefont {Y.}~\bibnamefont {Luo}}, \bibinfo
  {author} {\bibfnamefont {Z.}~\bibnamefont {Ren}}, \bibinfo {author}
  {\bibfnamefont {Z.}~\bibnamefont {Zhu}}, \bibinfo {author} {\bibfnamefont
  {C.}~\bibnamefont {Wang}}, \bibinfo {author} {\bibfnamefont {X.}~\bibnamefont
  {Xu}}, \bibinfo {author} {\bibfnamefont {Q.}~\bibnamefont {Tao}}, \bibinfo
  {author} {\bibfnamefont {G.}~\bibnamefont {Cao}}, \ and\ \bibinfo {author}
  {\bibfnamefont {Z.}~\bibnamefont {Xu}},\ }\href@noop {} {\bibfield  {journal}
  {\bibinfo  {journal} {New J. Phys.}\ }\textbf {\bibinfo {volume} {11}},\
  \bibinfo {pages} {025007} (\bibinfo {year} {2009})}\BibitemShut {NoStop}%
\bibitem [{\citenamefont {Weber}\ \emph {et~al.}(2006)\citenamefont {Weber},
  \citenamefont {Cosceev}, \citenamefont {Drobnik}, \citenamefont {Fai\ss{}t},
  \citenamefont {Grube}, \citenamefont {Nateprov}, \citenamefont {Pfleiderer},
  \citenamefont {Uhlarz},\ and\ \citenamefont {L\"ohneysen}}]{F.Weber}%
  \BibitemOpen
  \bibfield  {author} {\bibinfo {author} {\bibfnamefont {F.}~\bibnamefont
  {Weber}}, \bibinfo {author} {\bibfnamefont {A.}~\bibnamefont {Cosceev}},
  \bibinfo {author} {\bibfnamefont {S.}~\bibnamefont {Drobnik}}, \bibinfo
  {author} {\bibfnamefont {A.}~\bibnamefont {Fai\ss{}t}}, \bibinfo {author}
  {\bibfnamefont {K.}~\bibnamefont {Grube}}, \bibinfo {author} {\bibfnamefont
  {A.}~\bibnamefont {Nateprov}}, \bibinfo {author} {\bibfnamefont
  {C.}~\bibnamefont {Pfleiderer}}, \bibinfo {author} {\bibfnamefont
  {M.}~\bibnamefont {Uhlarz}}, \ and\ \bibinfo {author} {\bibfnamefont {H.~v.}\
  \bibnamefont {L\"ohneysen}},\ }\href {\doibase 10.1103/PhysRevB.73.014427}
  {\bibfield  {journal} {\bibinfo  {journal} {Phys. Rev. B}\ }\textbf {\bibinfo
  {volume} {73}},\ \bibinfo {pages} {014427} (\bibinfo {year}
  {2006})}\BibitemShut {NoStop}%
\bibitem [{\citenamefont {von Helmolt}\ \emph {et~al.}(1993)\citenamefont {von
  Helmolt}, \citenamefont {Wecker}, \citenamefont {Holzapfel}, \citenamefont
  {Schultz},\ and\ \citenamefont {Samwer}}]{R.Helmolt}%
  \BibitemOpen
  \bibfield  {author} {\bibinfo {author} {\bibfnamefont {R.}~\bibnamefont {von
  Helmolt}}, \bibinfo {author} {\bibfnamefont {J.}~\bibnamefont {Wecker}},
  \bibinfo {author} {\bibfnamefont {B.}~\bibnamefont {Holzapfel}}, \bibinfo
  {author} {\bibfnamefont {L.}~\bibnamefont {Schultz}}, \ and\ \bibinfo
  {author} {\bibfnamefont {K.}~\bibnamefont {Samwer}},\ }\href {\doibase
  10.1103/PhysRevLett.71.2331} {\bibfield  {journal} {\bibinfo  {journal}
  {Phys. Rev. Lett.}\ }\textbf {\bibinfo {volume} {71}},\ \bibinfo {pages}
  {2331} (\bibinfo {year} {1993})}\BibitemShut {NoStop}%
\end{thebibliography}%

\end{document}